\documentclass[aps,letterpaper,showpacs,preprint]{revtex4-1}

\usepackage{epsfig}
\usepackage{graphicx}
\usepackage{amsmath}
\usepackage{listings}% Include figure files
\usepackage[english]{babel}
\usepackage{amssymb}
\begin{document}

\title{Study of heat conduction  in systems with Kolmogorov-Arnold-Moser phase space structure}

\author{ I. F. Herrera-Gonz\'{a}lez${}^{1,2}$, H. I. P\'erez-Aguilar${}^{1}$, A. Mendoza-Su\'arez${}^{1}$, and E. S. Tututi${}^{1}$\footnote{tututi@umich.mx}}
\affiliation{$^{1}$Facultad de Ciencias F\'{i}sico-Matem\'{a}ticas, UMSNH, 
Av. Francisco J. M\'ujica S/N, 58060, Morelia, Michoac\'{a}n, M\'exico\\
$^{2}$Instituto de F\'{\i}sica, Benem\'erita Universidad Aut\'{o}noma de Puebla, 
Apartado Postal 72570, M\'exico}

\pacs{05.60.-k, 44.10.+i, 05.45.-a, 05.70.Ln}

%\date{\today}

\begin{abstract}
 We study heat conduction in a billiard channel formed by two sinusoidal walls and the diffusion of particles in the corresponding channel of infinite length;  the latter system has an infinite horizon,  i.e, a particle can travel an arbitrary distance without colliding with the rippled walls.  For  small ripple amplitudes, the dynamics of the heat carriers is regular and   analytical results for the temperature profile and heat flux are obtained using an effective potential. The study also proposes a formula for the temperature profile that is valid for any ripple amplitude. When the dynamics is regular, ballistic conductance and ballistic diffusion are present. The Poincar\'e plots of the associated dynamical system (the infinitely long channel) exhibit the generic transition to chaos as ripple amplitude  is increased. When no Kolmogorov-Arnold-Moser (KAM) curves are present to forbid the connection of all chaotic regions, the mean square displacement grows asymptotically with time $t$ as $t \log\left(t\right) $.  
\end{abstract}

\maketitle
%%%%%%%%%%%%%%%%%%%%%%%%%%%%%%%%%%%%%%%%%%%%%%%%%%%%%%%%%%%%%%%%%%%%%%%%%%%%%%%%%%%%%%%%%%%%%%%%%%%%%%%%
\section{Introduction}
Much attention has been paid in the past decade to the study of heat conduction in 1D and 2D systems via their underlying dynamics. The main question to be answered is how the Fourier law of heat conduction is related to the microscopic dynamics of the system. On the one hand, a convergent conductivity has been shown in the ding-a-ling model \cite{ding} (where oscillators exchange energy via intermediate hard spheres), which is chaotic, and  in the Lorentz channel (a channel with circular scatterers placed periodically) which is fully hyperbolic \cite{lorentz}. On the other hand, for the Fermi-Pasta-Ulam chain, where oscillators are coupled to non-linear terms, heat conductivity is  abnormal even above the chaotic threshold \cite{Fermi}; while the serpent billiard, a channel with parallel semi-circular walls, exhibits marginally normal diffusion even under conditions of  global chaos \cite{serpent}. Since normal diffusion is at the root of normal heat transport and abnormal diffusion leads to abnormal heat transport \cite{abnormal}, the latter system cannot obey Fourier's law. Thus, the  positivity of the Lyapunov exponent is neither a sufficient nor necessary condition for inducing normal transport properties, since there are billiard gas models (polygonal channels) with linear dynamical instability and yet they exhibit normal transport properties  \cite{tria,tria1,tria2,tria3,larralde}. In addition, it seems that for chaotic systems a strong degree of chaos, which translates into a global chaotic dynamics, is required to obtain normal transport properties.

Although the precise conditions behind the onset of Fourier's law remain unknown despite decades of intensive studies, many interesting properties of the heat mechanism have been discovered, including the possibility of controlling heat flux: for example, there are 1D chains that act as thermal rectifiers \cite{Casati,Casati3,Pereira}, billiard models where particle interactions (or an external magnetic field) produce thermal rectification \cite{Monasterio, Casati2} and graded systems in which rectification does not decay with system size \cite{Pereira1,Pereira2}. Moreover, a temporally alternating bath temperature can be used to generate a steady heat flow against a thermal bias \cite{Hangi2}, while an important increasing of the thermoelectric efficiency is reported in billiard-like systems \cite{Casati4}. Another interesting, related issue is the possibility of controlling the stochastic transport of particles (in the absence of a thermal gradient) which can be achieved in systems possessing spatial or  dynamical symmetry breaking with the aid of external unbiased input signals \cite{Hangi}.  

\begin{figure}[htp]
\begin{center}
\includegraphics[scale=0.3]{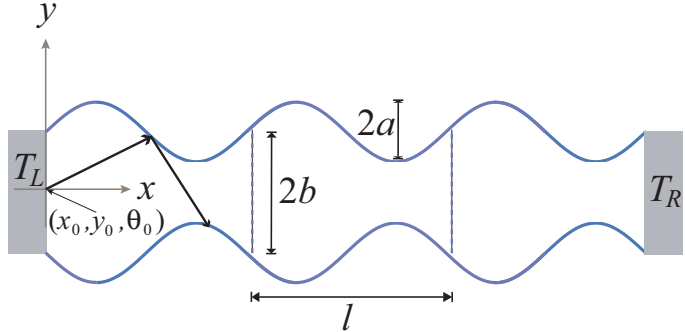}
\caption{(Color on line) Geometry of the channel in contact with two thermal baths.}
\label{fig6}
\end{center}
\end{figure}

Much attention has been paid to billiard systems when they exhibit strong chaos (global chaos), but few studies have focused on the relation between the degree of chaos in the system and their transport properties \cite{chaos,Luna}, despite the importance  of this issue, as in many systems chaotic and regular dynamics coexist. In addition, the system introduced in Ref. \cite{chaos} is a  peculiar chaotic system whose phase space does not have the typical KAM structure of generic Hamiltonian systems. While diffusion in systems (quasi-1D cosine billiard) with KAM phase space structure is studied in Ref. \cite{Luna}, but no connection with the thermal transport properties  is probed. Therefore, the main purpose of our study is to analyze the effect of the degree of chaos on the thermal transport properties of systems with KAM phase space structure.\\
In this paper, we consider a two-dimensional billiard model formed by two sinusoidal walls. The two ends of the channel are connected to Gaussian-type thermal baths (see Fig. \ref{fig6}) and the corresponding infinite length channel has an infinite horizon. The average width of the channel chosen is much smaller than its length in order to keep kinetic excitation in the transverse direction frozen. The study of thermal properties in this system is interesting because it exhibits Poincar\'e sections  with a KAM structure typical of generic Hamiltonian systems \cite{Ivan}. In addition, at small ripple amplitudes, we obtain an estimation of the temperature profile and heat flux by means of an effective potential. The result of our estimation  is corroborated by numerical calculations.\\
The paper is organized as follows. In Sec II we introduce the model and  discuss the dynamic properties of the system. Sec. III focuses on the study of the thermal transport properties of the channel when ripple amplitude is small. For this case, we use the effective potential mentioned in the previous section to obtain analytical results for the stationary heat flux and temperature profile. Section IV examines how the  degree of chaos affects the thermal transport properties in the channel, focusing on the point at which  the transition from regular dynamics to chaotic dynamics occurs. When the ripple amplitude is sufficiently large, the results are compared to those obtained for the Lorentz channel and for the system introduced in Ref. \cite{chaos}. Finally, concluding remarks are presented.
%%%%%%%%%%%%%%%%%%%%%%%%%%%%%%%%%%%%%%%%%%%%%%%%%%%%%%%%%%%%%%%%%%%%%%%%%%%%%%%%%%%%%%%%%%%%%%%%%%%%%%%%%%%%%%%%%%%%%%%%%
\section{System Model and Dynamics}
 We considered an infinite channel formed by two sinusoidal walls; the profiles of the upper and lower walls, respectively, are determined by
\begin{eqnarray}
y_1=b+a\sin(2\pi x) \qquad y_2=-b+a\sin(2\pi(x+r)),
\end{eqnarray}
where $a$ is the amplitude of the ripples, $r$ denotes the phase difference between the upper and lower walls, and $b$ is half of the average width of the channel. All quantities are rescaled to the length $l$ of one period. In this study, we consider the following fixed values: $b=\mbox{0.1}$ and $r=1/2$. When we take other  values for $r$, the main conclusions of this paper remain essentially  the same.

\begin{figure}[htp]
\includegraphics[scale=0.64]{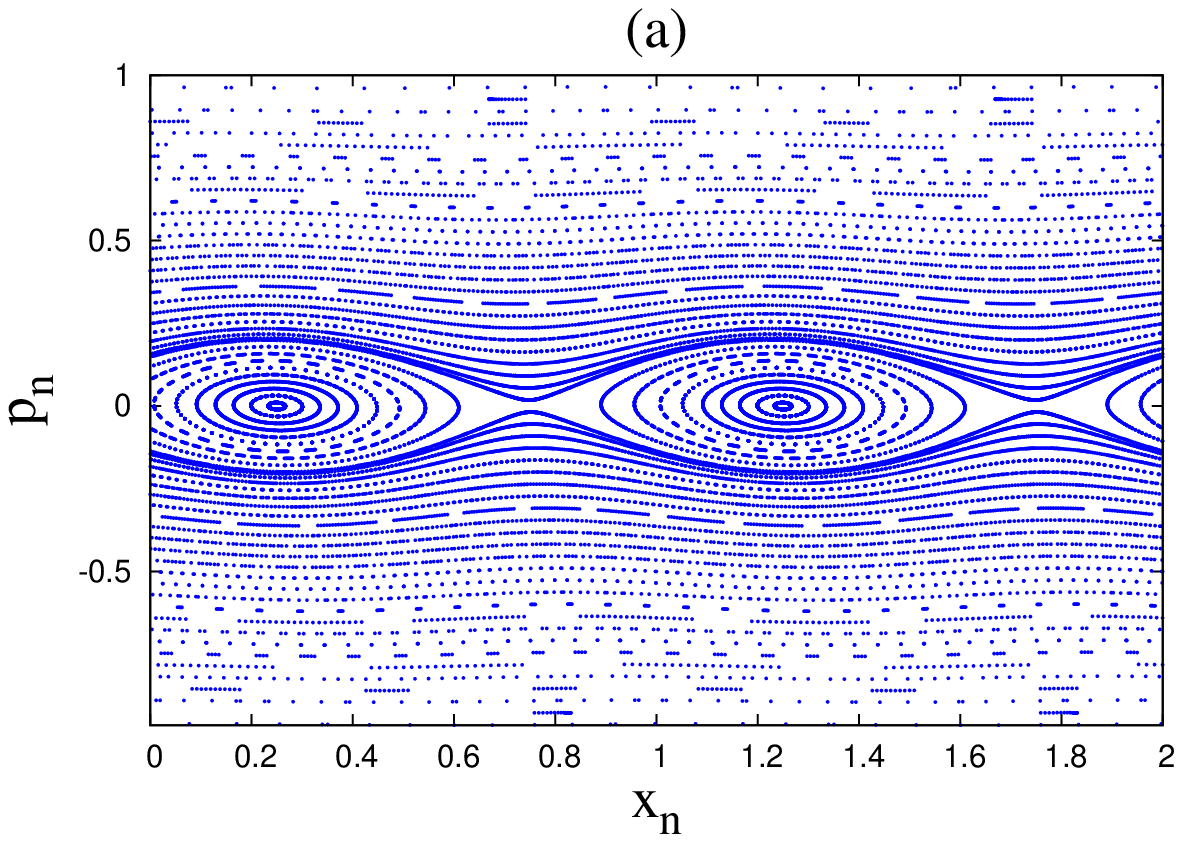}
\includegraphics[scale=0.64]{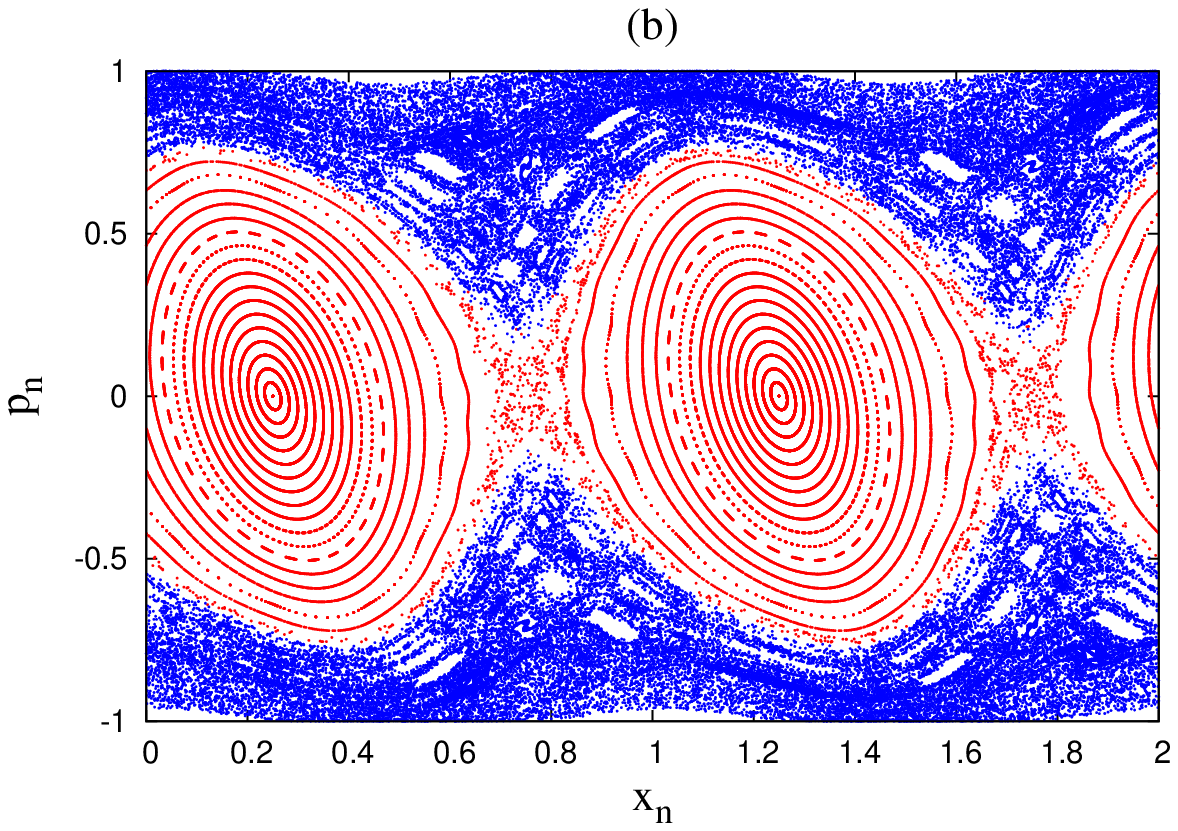}
\includegraphics[scale=0.64]{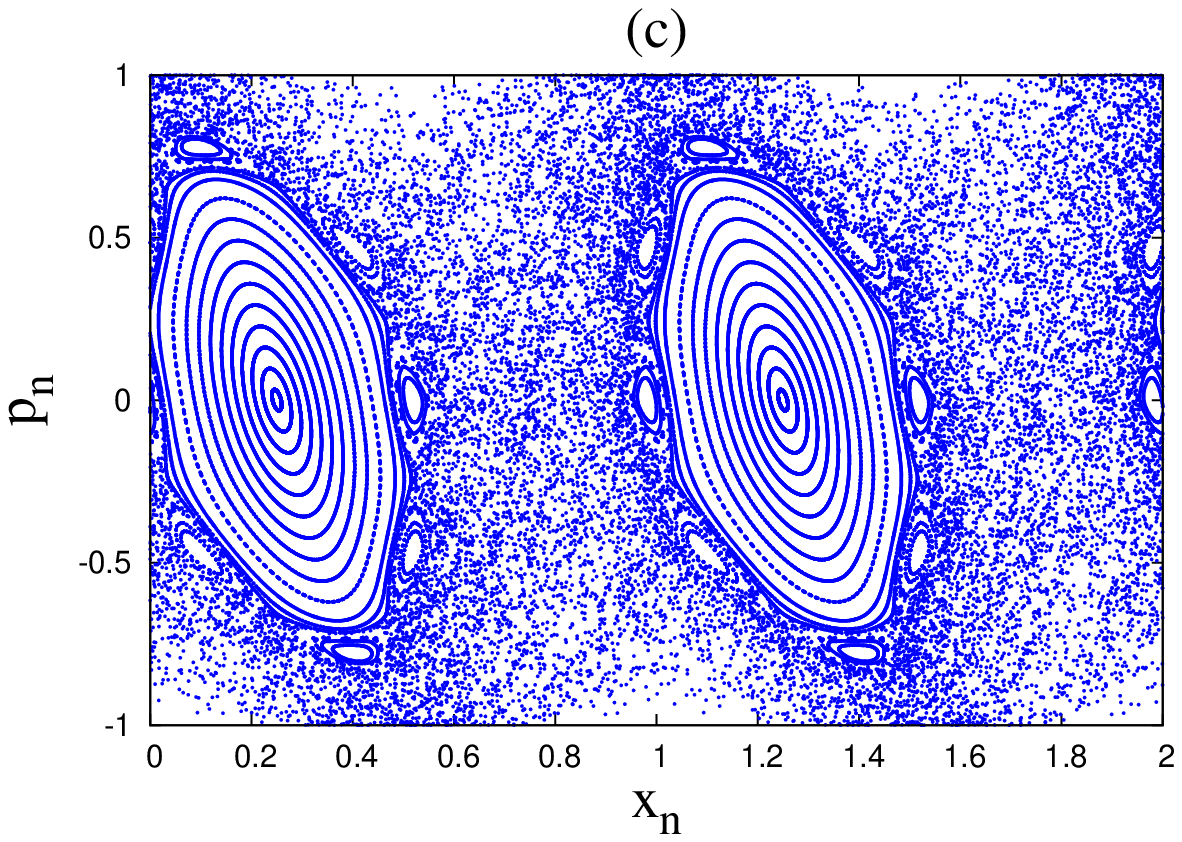}
\includegraphics[scale=0.64]{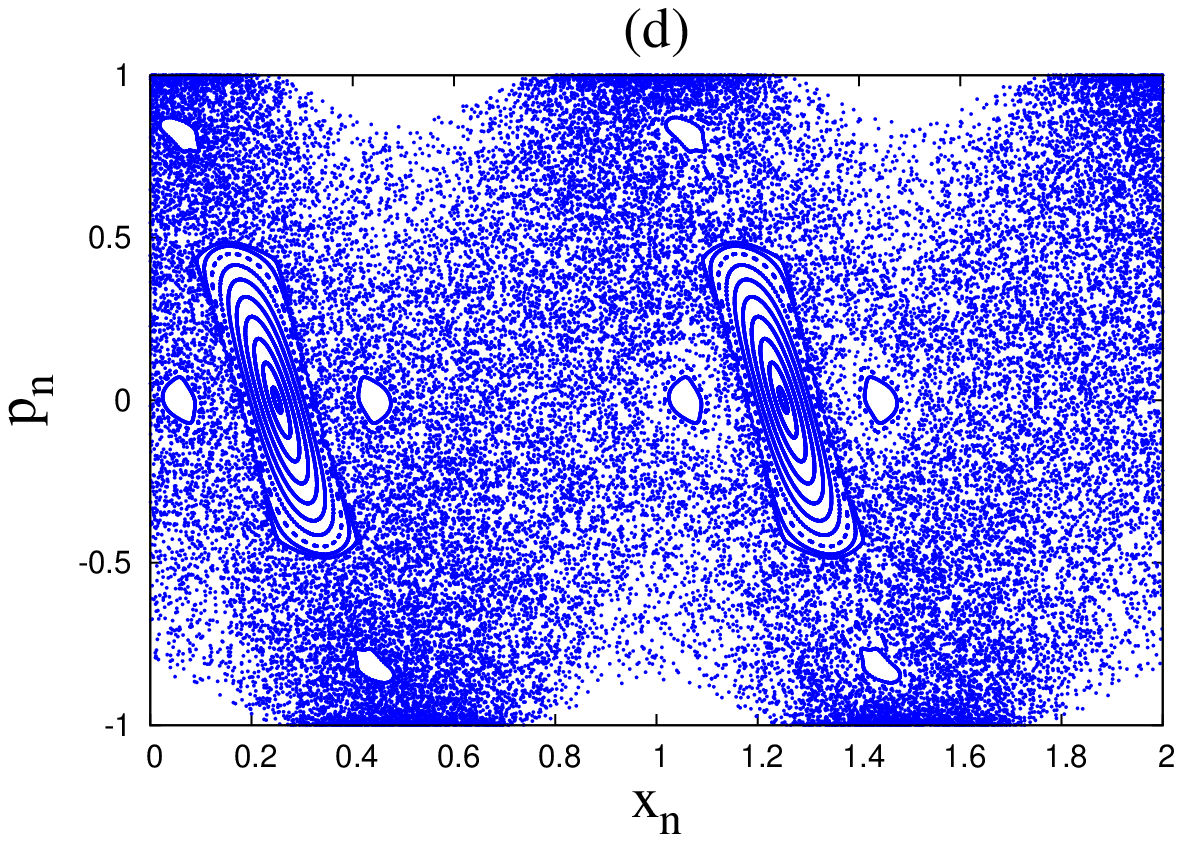}
\caption{\label{figu2}(Color on line) Poincar\'e plots for $r=1/2$ and $b=0.1$. (a) $a=0.001$, (b) $a=0.02$, (c) $a=0.04$, and (d) $a=0.09$.}
\end{figure}

 The dynamics of a particle that collides specularly with the rippled walls can be described qualitatively by the Poincar\'e  section $(x_n,p_n=\cos(\theta_n))$, where $x_n$ is  the $x$-coordinate of the particle  and $\theta_n$ the angle that the trajectory of the particle forms with the positive $x$-axis just after the $n$-th collision with any wall. To obtain all possible  orbits in the Poincar\'e section (PS),  several initial conditions $(x_0,y_0,\theta_0)$ must be taken into account. Here, $(x_0,y_0)$ is a point in the configuration space that corresponds to the initial $x$- and $y$-coordinates of the particle, while $\theta_0$ is the angle that the initial trajectory of the particle forms with the positive $x$-axis at the departure point $(x_0,y_0)$. Due to the periodicity of the channel, the structure of the PS is periodic; therefore, we may choose the $x$-domain in the interval $[0,2]$, which corresponds to two ripple periods, so as to obtain a complete panorama of the dynamics of the system. Poincar\'e plots for four different  values of $a$ are shown in Fig. \ref{figu2}. When the ripple amplitude is small, the dynamics of the system is regular and Poincar\'e sections resemble the phase space of a one-dimensional pendulum (see Fig. \ref{figu2}(a)). The elliptical orbits correspond to particles trapped in the channel that move adiabatically backwards and forwards around a stable fixed point. The trajectories outside the librational region  represent a particle traveling to the left ($p_n<0$) or right ($p_n>0$) of the channel (rotational motion). When $a$ is increased, the dynamics is still regular up to $a \lesssim \mbox{0.015}$, and the region of librational motion occupies a larger size. For the interval $\mbox{0.015} \lesssim a \lesssim \mbox{0.025}$ the separatrix becomes chaotic with some sizable width (see Fig. \ref{figu2}(b)) that increases as ripple amplitude becomes larger. There are two KAM curves (KAM barriers) that forbid the connection of all chaotic regions. They are found in the limits between the red region (light gray) and the blue region (dark gray) shown in Fig \ref{figu2}(b).  Therefore, motion is unidirectional for initial conditions lying outside both the librational region and the chaotic separatrix (we refer to this situation as unidirectional mixed chaos). There is a critical value of ripple amplitude ($a\approx \mbox{0.025}$) at which all chaotic regions are connected due to the destruction of the KAM curves. In this case, we found a first order resonant island surrounded by a chaotic sea (see Fig. \ref{figu2}(c)) and the reversal of the travel direction, for initial conditions falling outside the librational region, is now possible (we refer to this situation as bidirectional mixed chaos). As the ripple amplitude is  increased further, the size of the region of librational motion becomes smaller, as shown in Fig. \ref{figu2}(d).

For sufficiently small ripple amplitudes, a particle executing librational motion, or rotational motion whose corresponding orbit in the PS lies near the separatrix, will collide with the walls almost perpendicularly,  as Fig. \ref{figu2}(a) shows.  Under this circumstance, we can estimate  the  critical angle  $\theta_0=\theta_c$ for which the particle  executes the largest librational motion. In this case, the reversal of travel direction occurs at a point that is arbitrarily close to the nearest hyperbolic fixed point $x^h_f$ to the right of $x_0$ if $\cos \theta_c>0$. Since the average width of the channel is small, $\theta_c$ is virtually independent of the initial condition $y_0$ (this is corroborated by our numerical results), and we can set the departure point at $(x_0, y_2(x_0))$ to estimate $\theta_c$. Therefore, $(x_0, \cos \theta_c)$ is an initial condition in the PS. In order to estimate $\theta_c$, the condition of adiabatic invariance is used
\begin{eqnarray}
D(x_n)\left|\sin \theta_n \right|=D(x_m)\left| \sin \theta_m \right|,
\label{ada}
\end{eqnarray}
written for two arbitrary points $(x_n,\theta_n)$, ($x_m,\theta_m$), with $D(x_n)$ as the separation between the rippled walls at the point of collision $x_n$. Setting $x_n=x_0$, $\theta_n=\theta_c$, then $x_m\approx x^h_f$ and $\theta_m\approx \frac{\pi}{2}$. Upon expanding through a Taylor series expression (\ref{ada}), but maintaining the terms of order $a/b$ and $\beta^2_c$, where $\beta_c=\pi/2-\theta_c$, we obtain   
\begin{eqnarray}
\left(\beta_c\right)^2\approx \frac{a}{b}\left[ \sin\left(2\pi(x^h_f+r)\right)+\sin \left(2\pi x_0\right)-\sin \left(2\pi x^h_f\right)-\sin \left(2\pi (x_0+r)\right)\right].
\label{critical1}
\end{eqnarray}
Using the conservation of the energy and taking into account  the condition of adiabatic invariance, which implies that the quantity $C(x_n)=v_0D(x_n)|\sin\theta_n|$ remains constant, and with $v_0$ representing the speed of the particle, we obtain
\begin{equation}
\frac{1}{2} \left(\dot x\right)^2=E-V(x),
\label{potential}
\end{equation}
where $E$ is the energy of the particle (with unit mass) and $V(x)=\frac{1}{2}\left(\frac{C}{D(x)}\right)^2$; which can be interpreted as an effective potential that explains why the Poincar\'e plots for a small ripple amplitude resemble the phase space of a simple pendulum. The hyperbolic fixed points  can be computed by obtaining the maxima of $V(x)$. In our specific model ($r=1/2$),  $x^h_f=3/4$ if $x_0=0$ and $\cos\theta_c>0$. In this case, expression (\ref{critical1}) is reduced to
\begin{equation}
\beta_c\simeq \left(\frac{2a}{b}\right)^{\frac{1}{2}}.
\label{critical}
\end{equation}
The condition of adiabatic invariance and the critical angles have been used previously to arrive at estimates of the electronic transport properties in rippled channels when the Poincar\'e plots of the associate dynamical system exhibit a pendulum-like phase space \cite{Ivan,Luna2}.

%%%%%%%%%%%%%%%%%%%%%%%%%%%%%%%%%%%%%%%%%%%%%%%%%%%%%%%%%%%%%%%%%%%%%%%%%%%%%%%%%%%%%%%%%%%%%%%%%%%%%%%%%%%%%%%%%%%%%
\section{Thermal properties of the channel at small ripple amplitude}
This section considers a finite version of the channel discussed in the previous section.  This channel consists of $N$ replicas (a replica is a fundamental cell formed with the rippled  walls and  the vertical dashed lines, as shown in Fig. \ref{fig6}) of  length  $l=1$.  One end of the channel is at $x_L=0$, so the other one  is at $x_R=N$. To induce heat transport, the channel is placed between two heat baths at temperatures $T_{L}$ and  $T_{R}$ ($T_L>T_R$, see Fig. \ref{fig6}), modeled by stochastic kernels of Gaussian type by
\begin{eqnarray}
P(v_x)=\frac{\left|v_x \right|}{T}\mbox{exp}\left(-\frac{v^2_x}{2T} \right), \qquad P(v_y)=\frac{1}{\sqrt{2\pi T}}\mbox{exp}\left(-\frac{v^2_y}{2T}\right),
\label{distri}
\end{eqnarray}
where  $P(v_x)P(v_y)$ is the probability distribution of the  velocities for the particles emerging from the baths. The values of $N$ are such that $N\gg b$; hence, the number of particles that can cross the channel without colliding with the rippled walls is much lower than that of those which will have collisions.

Due to the fact that the particles do not interact among themselves, the motion of a single particle is considered over a long time period as it collides with the rippled walls and thermal baths. Collisions with the rippled boundaries are specular, and the velocity of the particle just after a collision with a thermal bath is determined by the distribution given by Eq. (\ref{distri}). Following the ideas from Ref. \cite{lorentz}, we divide the configuration space into slices $\{C_i\}$ (eventually a slice will be taken as a fundamental cell). The time that the particle spends  in the slice in the  $j$-th visit is denoted by $t_j$, and the total number of times that the particle crosses $C_i$ is $M$. The temperature of the slice $C_i$ is defined by
\begin{equation}
T_{C_i}=\frac{\sum^M_j t_jE_j(C_i)}{\sum^M_j t_j},
\label{tem}
\end{equation}
where $E_j(C_i)$ is the kinetic energy of the particle at the $j$-th visit of the slice $C_i$. The heat flux of one particle can be defined by
\begin{equation}
j_1(t_c)=\frac{1}{t_c}\sum^{N_c}_{k=1} (\triangle E)_k,
\label{flux}
\end{equation}
where $(\triangle E)_k$ is the energy change of the particle at the $k$-th collision with a thermal bath, and $t_c$ is the total time that the particle takes to collision $N_c$ times with a thermal bath.
The correct way to obtain the thermodynamic limit is to keep the number of particles per cell fixed as the size $N$ of the channel increases. For instance, if a single particle per cell is considered, then the heat flux in a channel of $N$ length is $j_N(t_c)=Nj_1(t_c)$. The thermal conductivity $\kappa$ is  determined by
\begin{equation}
\kappa=\frac{N^2 j_1(N)}{T_L-T_R} \sim N^2 j_1(N),
\label{conducty}
\end{equation}
where we have defined the thermal conductivity through the Fourier law of heat conduction. The next step is to obtain an expression of the temperature profile and heat flux in the stationary regime for the case of a small ripple amplitude. In the interests of simplicity, but without losing generality, the slices $\{C_i\}$ are assumed to be equal to the fundamental cells. For sufficiently small ripple amplitudes, the librational orbits, or rotational orbits near the separatrix, represent a particle bouncing off the walls almost perpendicularly.  Thus, the  time  in the $j$-th visit that the particle spends within the slide $C_1$ (which is in direct contact with the hot bath) for these rotational orbits is given, approximately, by                                  
\begin{eqnarray}
\nonumber
t^{(r)}_j&=&\sum_{C_1}\left| \frac{x^{(j)}_{k+1}-x^{(j)}_{k}}{\dot{x}^{(j)}_k}\right|\approx \int^{1}_0 \frac{dx}{\sqrt{E_j-V_j(x)}}\\
&=&\int^1_0 \frac{D(x)dx}{v^{(j)}_0\sqrt{D^2(x)-D^2(0)\sin^2(\theta^{(j)}_0)}},
\label{time}
\end{eqnarray}
while  for the librational orbits, we have
\begin{eqnarray}
\nonumber
t^{(l)}_j&=&\sum_{C_1} \left|\frac{x^{(j)}_{k+1}-x^{(j)}_{k}}{\dot{x}^{(j)}_k}\right|\approx 2\int^{x_c}_0 \frac{dx}{\sqrt{E_j-V_j(x)}}\\
&=&2\int^{x_c}_0 \frac{D(x)dx}{v^{(j)}_0\sqrt{D^2(x)-D^2(0)\sin^2(\theta^{(j)}_0)}}.
\label{time1}
\end{eqnarray}      
$\dot{x}^{(j)}_k(x^{(j)}_k)$ denotes the velocity in the $x$-direction just after the $k$-th collision with any wall (the $x$-coordinate of the particle at the $k$-th collision with any wall) during the $j$-th visit to cell $C_1$. $\dot{x}^{(j)}_k$ as a function of $x^{(j)}_k$ is determined using the effective potential given by Eq. (\ref{potential}), while the trajectory of the particle becomes known once  the random initial  conditions $v^{(j)}_0,\theta^{(j)}_0$ are determined just after a collision with a thermal bath. When the particle collides with a thermal bath again, we continue computing its trajectory once the new initial conditions $v^{(j+1)}_0,\theta^{(j+1)}_0$ are known. Here $\theta^{(j)}_0$, $v^{(j)}_0$, $E_j$, $V_j(x)$ have the same meaning as $\theta_0$, $v_0$, $E$, $V(x)$, introduced above, respectively. $D(x)$ is the distance between the  rippled walls at point $x$, which in our specific model is given by
\begin{eqnarray*}
D(x)=2\left(b-a\sin(2\pi x)\right).
\end{eqnarray*}
Expressions (\ref{time}) and (\ref{time1}) are obtained using Eq. (\ref{potential}), where we  set $C^{(j)}=D(0)v^{(j)}_0|\sin \theta^{(j)}_0|$. The factor 2 in Eq. (\ref{time1}) derives from the fact that the time that the particle takes to execute librational motion before being reinjected into the same bath is twice the time it spends going from the thermal bath to the returning point $x_c$, where the particle changes its direction of travel and the corresponding $x$-velocity is approximately zero. Then, $x_c$  is determined by the nearest positive root of $E-V(x)=0$ to $x_0=0$. In the case of the slice $C_N$, we find similar expressions, but $x_c$ is now the nearest  root to the left of $x_0=N$ (note that, due to the system periodicity, this  is equivalent to take the absolute value of the nearest negative root to $x_{0}=0$) that we denoted as $x_{cN}$. Particles emerging from the thermal baths cannot have access to librational orbits in $C_i$ cells with $i=2,3,\dots N-1$; thus the motion in these cells is rotational.  

Rotational orbits that do not lie near the separatrix are almost flat as shown in Fig 2(a); so  $t^{(r)}_j\approx (v^{(j)}_0\cos \theta^{(j)}_0)^{-1}$. A rotational orbit near the separatrix means that its corresponding initial condition satisfies $\cos \theta_c\simeq \left(a/2b\right)^{\frac{1}{2}} \lesssim \cos \theta^{(j)}_0$ . For rotational orbits that do not lie near the separatrix, $t^{(r)}_j$ can also be approximated by Eq. (\ref{time}), since upon 
 developing it in power series of $a/b$, we obtain $t^{(r)}_j=(v^{(j)}_0\cos \theta^{(j)}_0)^{-1}+O\left(a^2/(b^2 \cos^4 \theta^{(j)}_0)\right)+O\left(a^2/(b^2 \cos^2 \theta^{(j)}_0) \right)$. Due to the periodicity of rotational motion, the time that the particle takes to cross any cell while executing librational motion, is approximated  by Eq. (\ref{time}). 

Equation (\ref{time})  can be developed in power series of $\left(a/(b\cos^2 \theta^{(j)}_0)+a^2/(b^2\cos^2 \theta^{(j)}_0)\right)$, but we decide not to present the expansion because as we approached an initial condition that falls near the separatrix, it became necessary to include more terms of the expansion; while for an initial condition that falls exactly on the separatrix all terms of the expansion must be included. When we consider librational motion, another expansion is necessary; so we found it more convenient  to keep Eqs. (\ref{time}) and (\ref{time1}) in their integral form.

Now we can establish the behavior of $t_j$ as a function of both the cell number and of $t^{(r)}_j$ and $t^{(l)}_j$, which is given by
\begin{equation}
\label{time2}
   t_j(C_i)   =
       \left\{
               \begin{array}{cc}
                t^{(r)}_j, & \mbox{if} \quad \theta_0< \theta_c \quad i=1,\dots N  \\
                t^{(l)}_j , & \mbox{if} \quad \theta_0\geq \theta_c \quad i=1,N \\
               \end{array}
        \right.,
\end{equation}  
where $\theta_c$ is the critical angle defined in Eq. (\ref{critical}). Expression (\ref{time2}) is derived using the arguments that follow. The particles that come from the thermal baths at an angle $\theta_0 \geq \theta_c$ execute librational motion in the cells $i=1$ or $i=N$, depending on the thermal bath from which they come. Those particles are then reinjected into the same heat reservoir, then $t^{(l)}_j$ is only defined in the cells $i=1,N$. $\theta_c$ is the same for particles coming from any heat bath because the spatial separation of the baths is equal to an integer number. When $\theta< \theta_c$,  particle motion is rotational and therefore periodic; thus the time required for a particle to cross cell $C_i$ is independent of the $i$ index.  

\begin{figure}[htp]
\includegraphics[width=7.9cm,height=6.9cm]{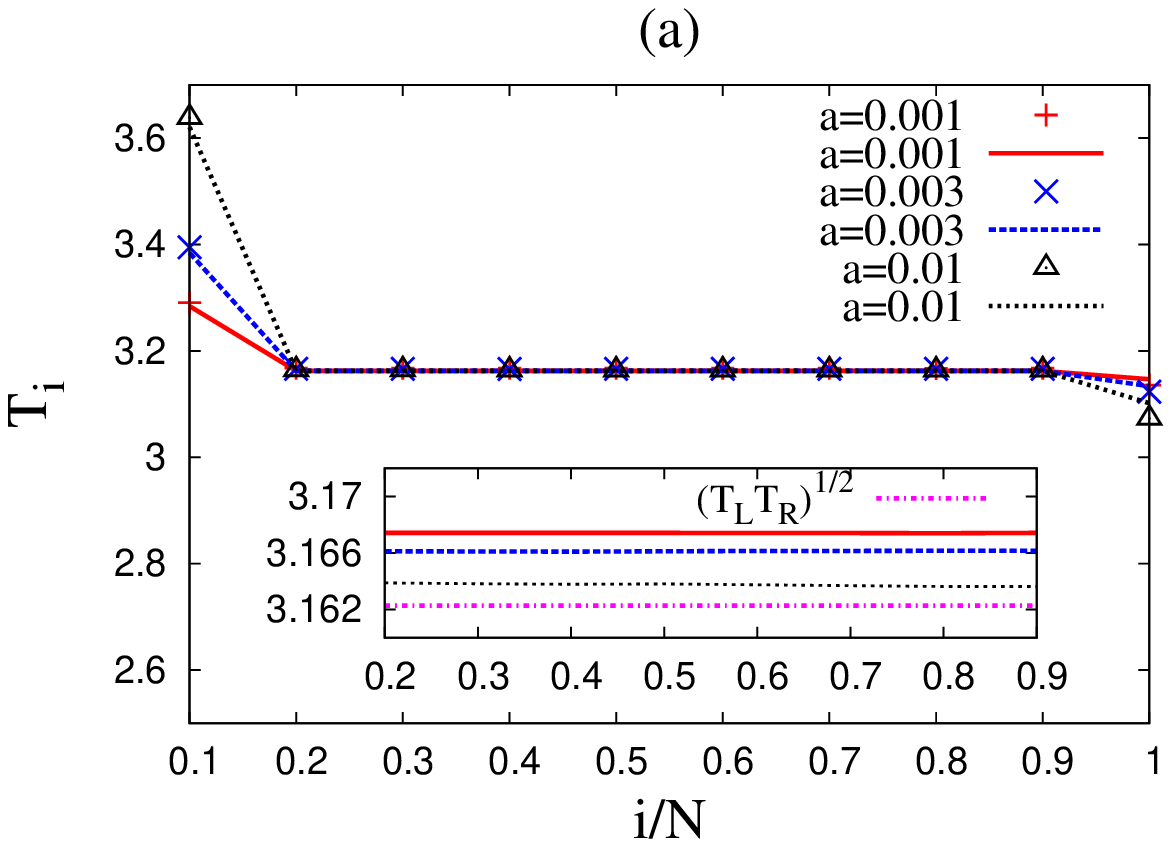}
\includegraphics[width=7.9cm,height=6.9cm]{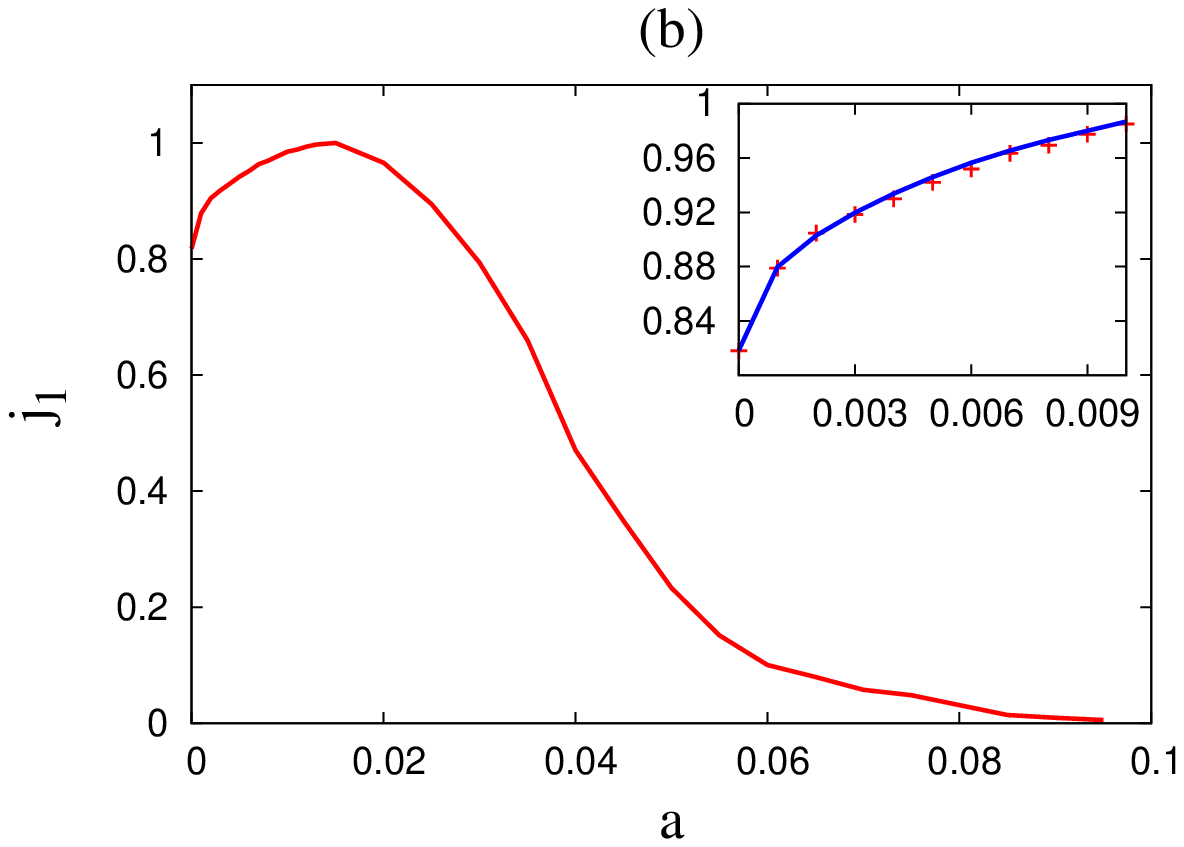}
\caption{\label{figu3} (Color on line) (a) Temperature profile for three different amplitudes; $a=\mbox{0.001}$, $a=\mbox{0.003}$ and $a=\mbox{0.01}$. Continuous lines represent the theoretical prediction (\ref{t1})-(\ref{t3}), while the different styles of dots represent the numerical data. The inset presents a zoom of the numerical temperature profile in the central part of the channel to show the accuracy of the theoretical prediction (\ref{t1}). (b) Heat flux of one particle as a function of ripple amplitude. The inset uses dots to show the numerical data, while the continuous line represents the theoretical prediction for one particle heat flux for the amplitude interval $[0,\mbox{0.01}]$. In both figures, $T_L=10$, $T_R=1$, and channel length is $N=10$.}
\end{figure}

To compute the temperature profile, we divide the particles in two types: those that come from the bath at temperature $T_{L}$ and those that come from the  bath at temperature $T_{R}$ at the moment they cross the slice $C_i$. In this way, definition (\ref{tem}) reads
\begin{equation}
T(C_i)=\frac{\left<tE\right>^{(i)}_L+\left<tE\right>^{(i)}_R}{\left<t\right>^{(i)}_L+\left<t\right>^{(i)}_R},
\label{prom1}
\end{equation}
where $\left<\cdots \right>^{(i)}_L$ denotes the time average at cell $C_i$ using probability distribution (\ref{distri}) with $T=T_L$; $\left<\cdots \right>^{(i)}_R$ has the same notation, but with $T=T_R$. By expressing the probability distribution (\ref{distri}) in polar coordinates $(v_0,\theta_0)$, and substituting Eq. (\ref{time2}) into Eq. (\ref{prom1}), we have
\begin{eqnarray}
\label{t1}
T(C_i)&\simeq&\sqrt{T_LT_R} \quad \mbox{for} \quad i=2,3,\dots,N-1, \\
T(C_1)&\simeq&\sqrt{T_LT_R}\frac{\sqrt{T_L}I_1(\theta_c)+\left(\sqrt{T_L}+\sqrt{T_R}\right)I_2(\theta_c)}{\sqrt{T_R}I_1(\theta_c)+\left(\sqrt{T_L}+\sqrt{T_R}\right)I_2(\theta_c)},\\
\label{t2}
T(C_N)&\simeq&\sqrt{T_LT_R}\frac{\sqrt{T_R}I_3(\theta_c)+\left(\sqrt{T_L}+\sqrt{T_R}\right)I_2(\theta_c)}{\sqrt{T_L}I_3(\theta_c)+\left(\sqrt{T_L}+\sqrt{T_R}\right)I_2(\theta_c)}.
\label{t3}
\end{eqnarray}
Here,
\begin{eqnarray*}
I_{1,(3)}(\theta_c)&=&2\int^{\beta_c}_0\int^{x_c(\beta_0),(x_{cN}(\beta_0))}_0 \frac{D(x)\sin(\beta_0)dxd\beta_0}{\sqrt{D^2(x)-D^2(0)\cos^2(\beta_0)}},\\
I_2(\theta_c)&=&\int^{\theta_c}_0\int^{1}_0 \frac{D(x)\cos(\theta_0)dxd\theta_0 }{\sqrt{D^2(x)-D^2(0)\sin^2(\theta_0)}},
\end{eqnarray*}
with $\beta_0=\frac{\pi}{2}-\theta_0$. The numerical results support our theoretical results for the temperature profile predicted by Eqs. (\ref{t1})-(\ref{t3}), as shown in Fig. \ref{figu3}(a), where it is clear that the temperature of the cells that are not in direct contact with the thermal baths is the geometric average of the temperatures of those baths. Deviations from this flat profile occur at the cells that are in direct contact with the heat reservoirs due to the librational motion of the particles in those cells. The theoretical predictions for these deviations agree with the numerical simulation. In the case of a flat channel ($a=0$) $\theta_c=\frac{\pi}{2}$ and the temperature profile is completely flat with value $\sqrt{T_LT_R}$, this result is exact and can be corroborated directly using definition (\ref{prom1}).  

Using a similar procedure that made it possible to obtain analytical results for the temperature profile, we obtain the following expression for heat flux  
\begin{eqnarray}
\nonumber
j_N(N)&=&\frac{\frac{1}{2}N\left(\left<v_0^2 \right>_L-\left<v_0^2 \right>_R\right)}{\left<t^{(l)}\right>_L+\left<t^{(l)}\right>_R+N\left(\left<t^{(r)}\right>_L+\left<t^{(r)}\right>_R \right)}\\
&\simeq&\frac{3N\sqrt{\pi}\sin \theta_c \left(T_L^{\frac{3}{2}}\sqrt{T_R}-T_R^{\frac{3}{2}}\sqrt{T_L}\right)}{2^{\frac{3}{2}}\left(\sqrt{T_R}I_1(\theta_c)+\sqrt{T_L}I_3(\theta_c)+N\left(\sqrt{T_L}+\sqrt{T_R} \right)I_2(\theta_c)\right)},
\label{flux1}
\end{eqnarray}
where the  resulting factor $N$ in the denominator comes from the periodicity of the rotational motion, and from the fact that a particle executing this kind of motion must travel through $N$ cells of unit length before colliding with the other thermal bath. At the   limit $N\gg1$, heat flux becomes independent of system size and  thermal conductivity scales as $\kappa \sim N$. We have corroborated numerically that the latter scaling behavior is valid as long as the
system exhibits regular dynamics. In the latter case, ballistic diffusion is also present.

Our theoretical prediction (\ref{flux1}) is confirmed by the numerical results, as the inset in Fig. \ref{figu3}(b) shows. We should point out an unexpected behavior of increasing heat flux when  $a$ is increased to a maximum value at $a\approx \mbox{0.015}$, when chaotic behavior appears.  This behavior can be explained as follows: the particles executing librational motion do not contribute to heat flux because they never reach the other side of the channel. This fact is associated with the term $\sin \theta_c$ in the numerator of Eq. (\ref{flux1}), which decreases as ripple amplitude is increased. However, the average time that the particle takes to travel from one thermal bath to the other becomes smaller as ripple amplitude increases,  because only  particles with  larger velocity components in the $x$-direction can reach the other side of the channel. The reduction of the  average time is associated with the term $I_2$ in the denominator of Eq. (\ref{flux1}); thus, we have  two competing mechanisms and the strongest one is that associated with the term $I_2$ when $N\gg 1$.

The behavior of the temperature profile predicted by Eqs. (\ref{t1})-(\ref{t2}) is similar to the corresponding one for a homogeneous harmonic chain between two Langevin heat baths \cite{lebo}; while the scaling behavior of the thermal conductivity of the homogeneous chain and our model is identical. The main difference between the two systems  comes from the fact that the bulk temperature of the harmonic chain is the arithmetical average of the temperatures of the two heat reservoirs. However, in both systems, deviations from a flat profile occur in those parts of the systems that are in direct contact with the reservoirs.  Another system that presents the same size-scaling behavior in heat flux and thermal conductivity is the harmonic chain, with specific long-range correlation of the isotopic disorder, attached to Langevin heat baths \cite{ivan2}.

\section{Larger ripple amplitude}
We consider four channels, with different values of $a$ that produce  different degrees of chaos in the system, in order to study their thermal transport properties. In case I, $a=0.02$ and unidirectional mixed chaos is present (see Fig. \ref{figu2}(b)). Case II corresponds to  $a=0.04$, where the system exhibits bidirectional mixed chaos (Fig. \ref{figu2}(c)). Cases III and IV correspond to $a=0.07$ (not shown) and $a=0.09$ (Fig. \ref{figu2}(d)), respectively; bidirectional mixed chaos is presented as well, but with a smaller region of librational motion  due to stronger chaotic behavior.

\begin{figure}[htp]
\includegraphics[width=7.9cm,height=6.9cm]{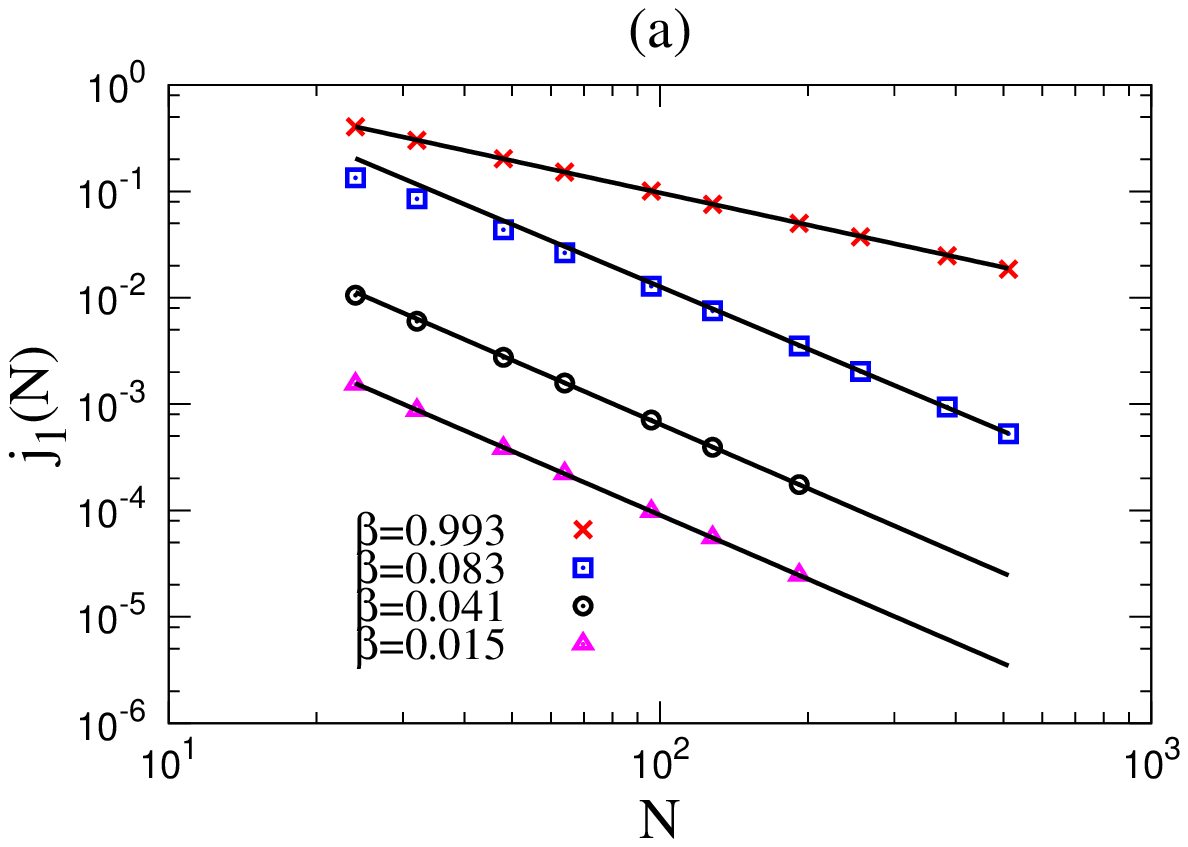}
\includegraphics[width=7.9cm,height=6.9cm]{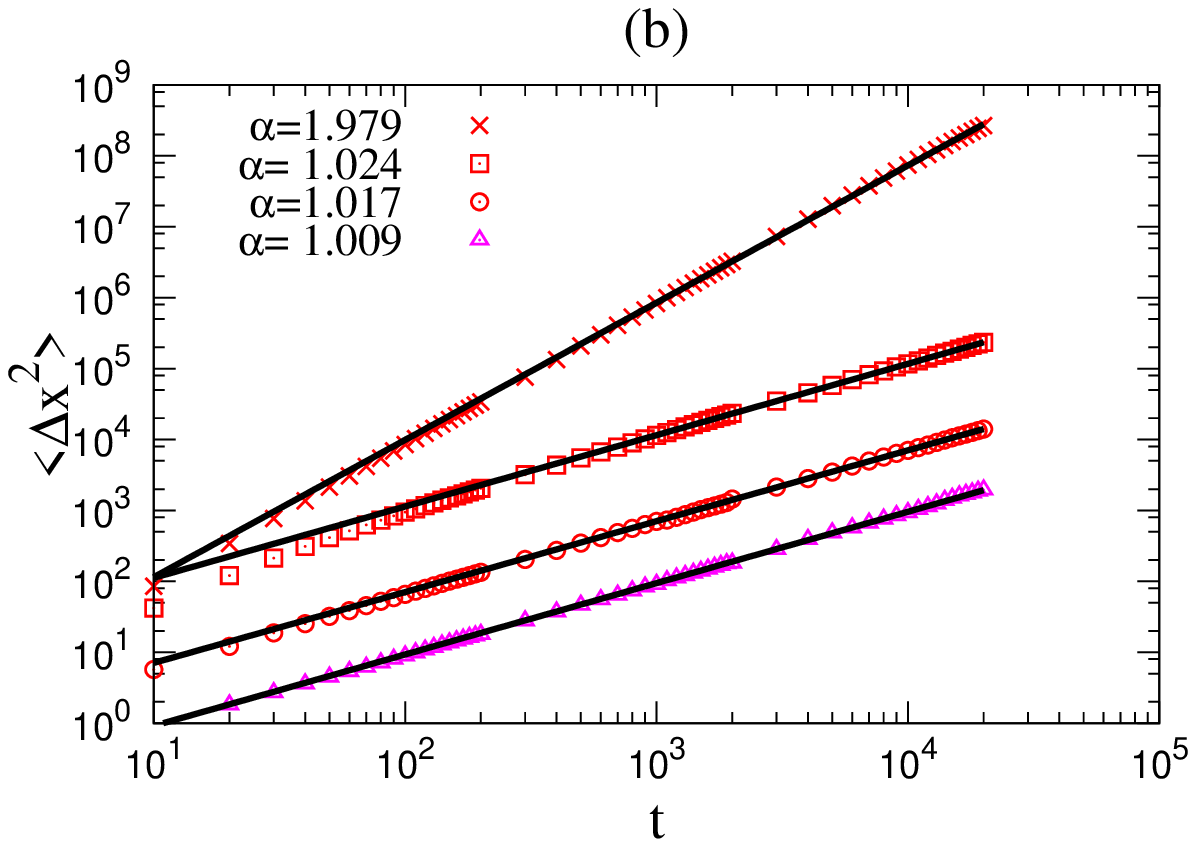}
\caption{\label{figu4} (a) (Color on line) Heat flux of one particle as a function of system size $N$ in log-log scale. (b) Mean square displacement $\left<\Delta x^2(t)\right>$ against time $t$. $10^5$ particles are used to obtain the diffusive properties of the system. Particles are initially at $x=0$, initial velocities obey the Maxwell-Boltzmann distribution at temperature $T=1$. In both figures, case I is ($\times$), case II ($\square$), case III ($\bigcirc$) and case IV ($\bigtriangleup$). }
\end{figure}

We are interested in studying two different, but related quantities: first, the scaling behavior of thermal conductivity in relation to the  system size; and, second, the mean square displacement $\left<\Delta x^2\right>=\left< \left(x(t)-x(0)\right)^2\right>$, where $\left<\dots \right>$ denotes an average over different initial conditions and $x(t)$ is the $x$-coordinate of the particle position at time $t$. For our numerical simulations of heat conduction, we choose $T_L=10$ and $T_R=1$. It is important to note that the  scaling of $j_1(N)$ with system size  does not depend on the properties of the thermal baths, but exclusively on the geometry of the channel \cite{tria}. Thus, the thermal conductivity defined by Eq. (\ref{conducty}) gives the same scaling behavior regardless of the temperature difference between the reservoirs. At the channel lengths explored in the numerical simulation, the thermal conductivity can be approximated by $\kappa= AN^{\beta}$ ($A$ and $\beta$ constants) for sufficiently long channels (see Fig. \ref{figu4}(a)). In the case of unidirectional mixed chaos, the exponent $\beta$ is close to $1$; ballistic behavior as reported  in Ref. \cite{Luna} is not present due to the presence of particles whose initial conditions fall in the chaotic separatrix and therefore their motion is not unidirectional. As soon as bidirectional mixed chaos appears, an abrupt decay occurs in  exponent $\beta$ from $\beta \approx 1$ to $\beta \approx 0$, because there are no more  particles with unidirectional motion.  As  ripple amplitude is increased, the exponent $\beta$ approaches zero. The diffusion of particles in the corresponding  channel of infinite length is also studied, and within the time interval explored in Fig. \ref{figu4}(b), the growth of the mean square displacement can be approximated by $Dt^{\alpha}$, for a sufficiently long time period. In reference to the scaling behavior of the thermal conductivity and  mean square displacement, we say "can be approximated", and not "scale as", because exponent $\beta$ exhibits a small change in relation to system size, and exponent $\alpha$ presents a small change in relation to time $t$. This means that these two exponents remain practically constant for  relatively long channel interval lengths and long time periods, respectively. The origin of the size-dependent and time-dependent exponents is discussed in the following paragraph. The numerical simulations of particle diffusion concord with the corresponding simulation of heat flux, in that exponents $\beta$ and $\alpha$ (see Figs. \ref{figu4}(a) and \ref{figu4}(b))  satisfy the relation $\beta=2-2/\alpha$, derived in Ref. \cite{abnormal}. It is interesting to note that for bidirectional mixed chaos, exponent $\alpha$ is close to 1 and exponent $\beta$ is close to 0.

\begin{figure}[htp]
\includegraphics[scale=0.7]{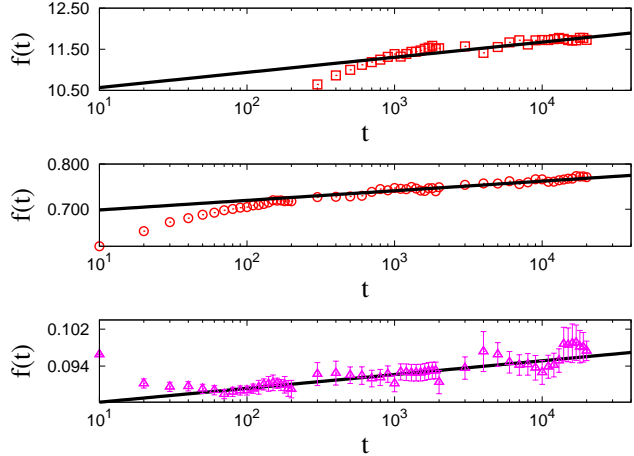}
\caption{\label{figu11}  (Color on line) $f(t)=\left(\left<\Delta x^2\right>/t\right)$ as a function of $\log\left(t\right)$. Figures from top to bottom correspond to cases II, III and IV, respectively. Solid lines represent the best fit of the numerical data. The fit gives the following values: case II, $c=0.18 \pm  1.9 \times 10^{-2}$, $d=10.00 \pm 0.2$; case III, $c=9.24\times 10^{-3}\pm 1.8\times 10^{-3}$, $d=0.67\pm 1.4 \times 10^{-2}$; case IV, $c=6.50 \times 10^{-3}\pm 3.47 \times 10^{-4}$, $d=8.45 \times 10^{-2} \pm 1.3 \times 10^{-4}$. In cases II and III, the statistical errors are less than the size of the points; in case IV, statistical errors are represented by the error bars.}
\end{figure}

In any model with an infinite horizon, there are particles that can travel arbitrary distances without suffering any collision with the boundaries at time $t$. These particles make a contribution to the velocity auto-correlation function that scales as $1/t$, and is the leading term if the corresponding contribution  of the rest of the particles ({\it i.e.}, those that do collide at time $t$) decays faster than $1/t$ \cite{Fried}. The algebraic decay $1/t$ of the velocity auto-correlation function has been confirmed numerically for the Lorentz gas with an infinite horizon \cite{Ble,Artuso,garrido}, and for a polygonal channel with an infinite horizon \cite{larralde}. Therefore, in our model, the mean square displacement must grow as $c t\log\left(t\right) +d t$ (the diffusion constant is related to the velocity auto-correlation function by the Einstein-Green-Kubo formula)  if the degree of chaos is strong enough; such a correction will appear in Fig. \ref{figu4}(b) as a small change in the $\alpha$ exponent and, therefore, a curvature in the mean square deviation will be visible for longer times. To observe the logarithmic correction, we plot $\left(\left<\Delta x^2\right>\right/t)$ against $\log\left(t\right)$, as shown in Fig. \ref{figu11}, and found that for cases II, III and IV, $\left< \Delta x^2 \right>=c t\log\left(t\right)+dt$ with  $0\lesssim c$ and $d \gg c$ (in general, we find this behavior for bidirectional mixed chaos). For case I, the scaling law $c t\log \left(t\right) +d t$ is incorrect; in fact, we do not know the correct asymptotic scaling law for unidirectional mixed chaos. However, $\left< \Delta x^2 \right>=Dt^{\alpha}$ (with $\alpha$ and $D$ constants) is a good approximation at the time period explored in the numerical simulations. The ratio $c/d$ decreases as ripple amplitude is increased, this is simply a consequence of the reduction of the number of particles that can travel arbitrarily far without colliding with the walls, as ripple amplitude becomes larger. In this sense, we can say that diffusion approaches normal behavior as ripple amplitude increases; hence, the same statement is also valid for heat flux. The marginally anomalous diffusion exhibited by our system (in the case of bidirectional chaos) provides  evidence that the contribution to the  velocity auto-correlation function from particles having collisions at time $t$ decays faster than $1/t$. Then, if the particles that can travel arbitrarily far without interacting with the walls were not present, diffusion should be normal. This argument indicates that normal diffusion can take place in systems with mixed phase space when there are not KAM curves that preclude connection of all chaotic regions, and when non-chaotic trajectories do not contribute to heat transport.
 
It must be stressed that for cases III and IV, all initial conditions of the particles coming from the thermal baths fall on the chaotic sea. However, if we place the thermal baths such that the initial conditions of these particles can have access to periodic and quasi-periodic orbits, exponents $\beta$ and $\alpha$ should be insensitive to this change because particles executing periodic or quasi-periodic motion do not contribute to heat flux for they are reinjected into the same bath from which they originally came. We have tested  the latter statement numerically by placing the thermal baths at $x_L=\mbox{0.25}$  and $x_R=\mbox{0.25}+N$. Here, particles have access to regular motion for all different cases. We find that exponents $\beta$ and $\alpha$  are practically the same than the corresponding ones given by Fig. \ref{figu4}. We report the following values for exponent $\beta$: case I, $\beta= \mbox{0.995}$; case II, $\beta= \mbox{0.089}$; case III, $\beta= \mbox{0.044}$, and case IV, $\beta=0.023$. 

\begin{figure}[htp]
\includegraphics[scale=0.64]{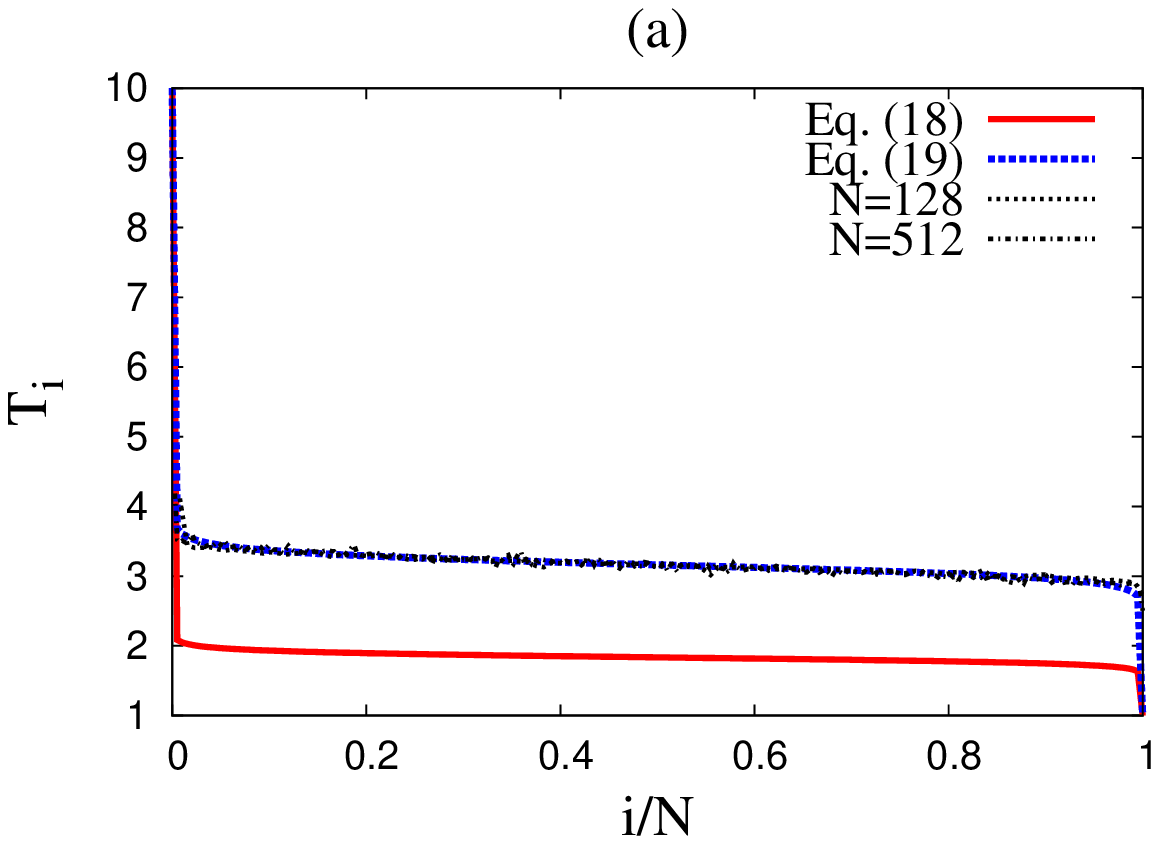}
\includegraphics[scale=0.64]{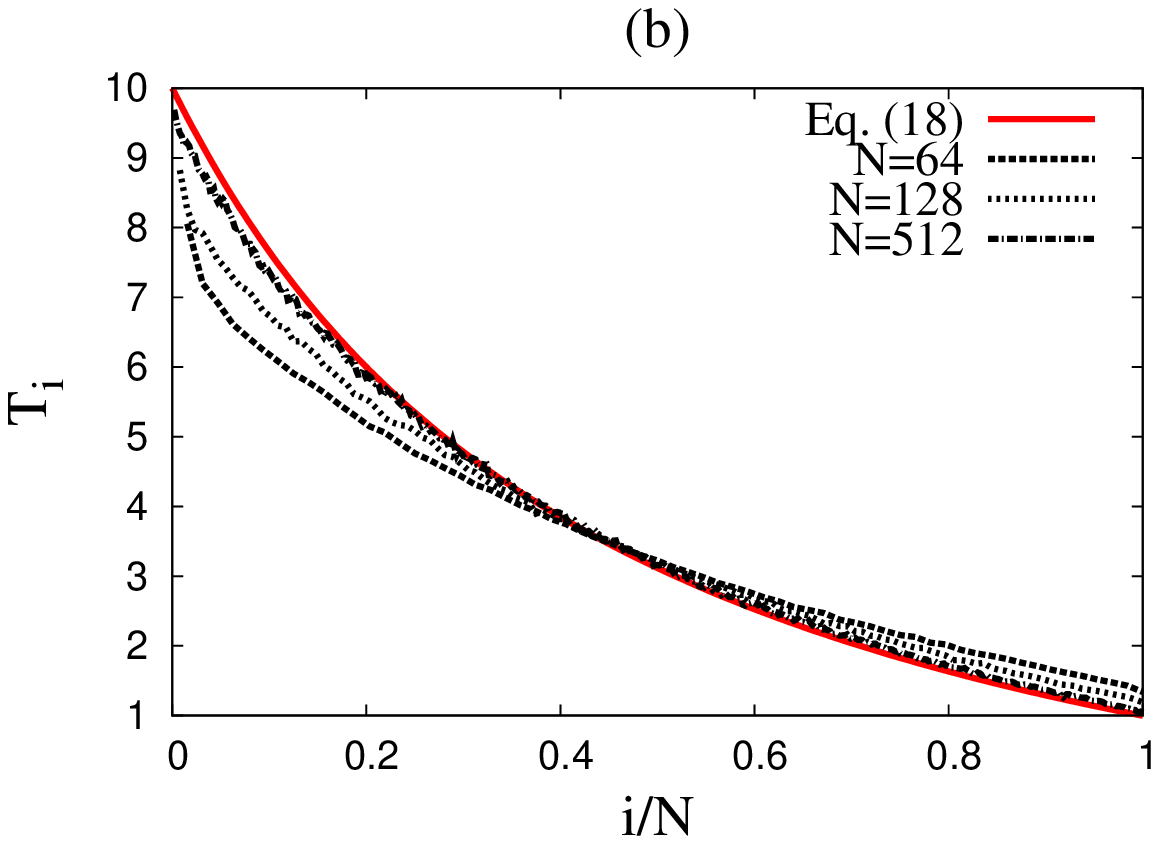}
\includegraphics[scale=0.64]{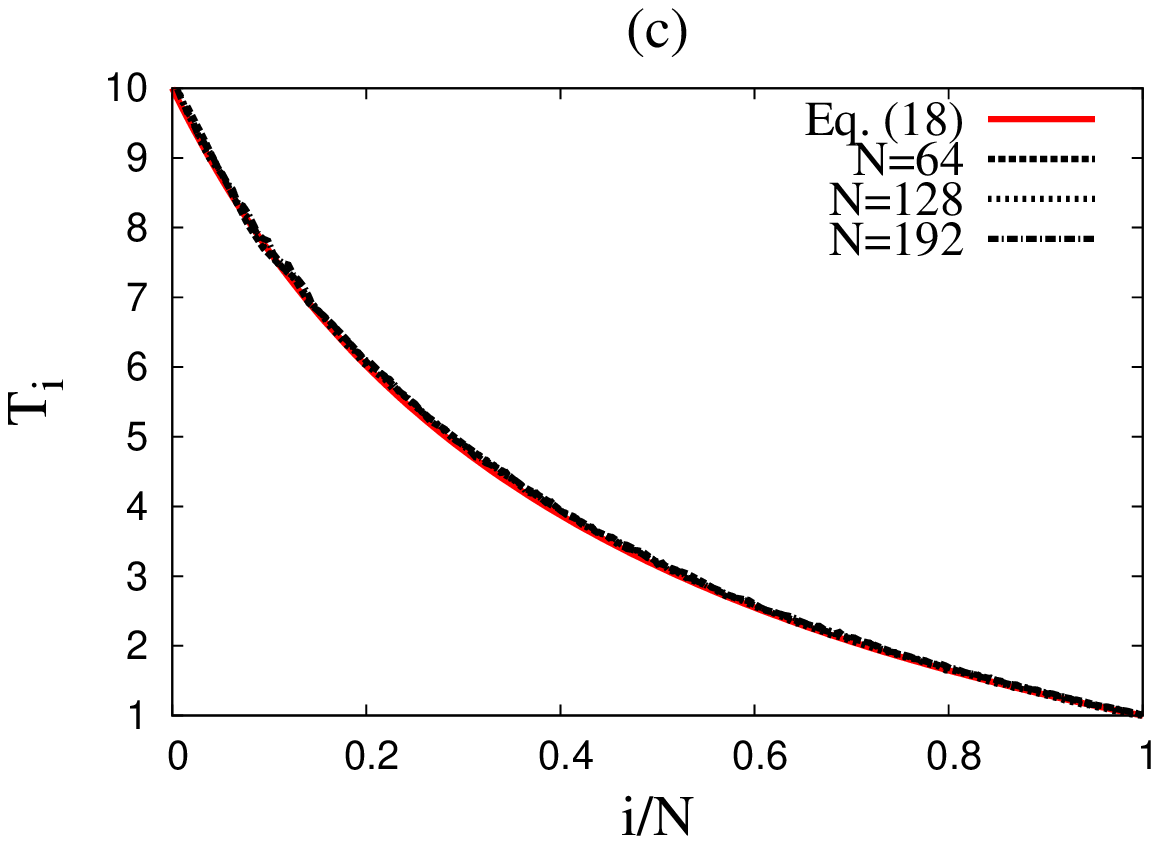}
\includegraphics[scale=0.64]{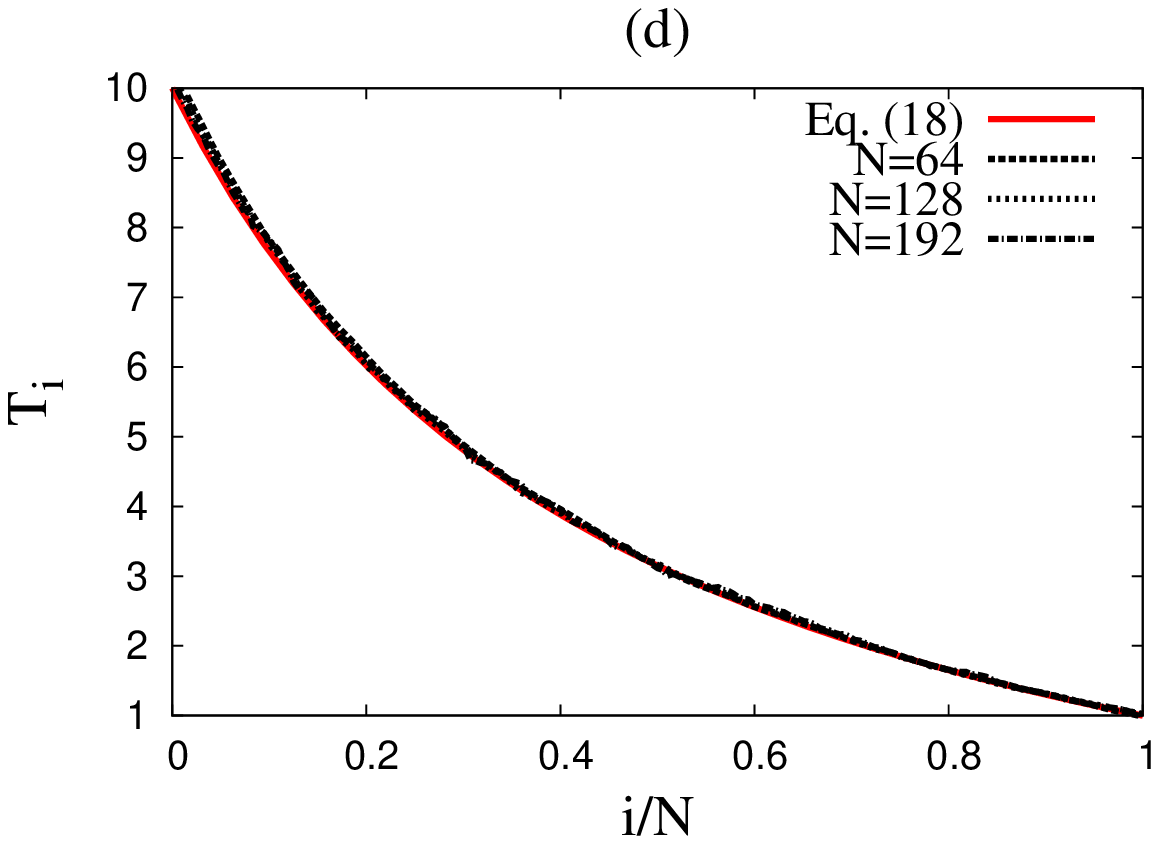}
\caption{\label{figu5} (Color on line) Internal local temperature as a function of the rescaled cell number $i/N$ with $T_L=10$ and $T_R=1$. (a) Corresponds to case I; (b) to case II; (c) to case IV; and,  (d) also to case IV, but with the thermal baths placed at $x_L=\mbox{0.25}$ and $x_R=\mbox{0.25}+N$. The solid lines in panels (c) and (d) correspond to the best fits for the numerical temperature profile at $N = 192$ with Eq. (\ref{ant}), the fit results are the following: panel (c), $\alpha=1.004$; panel (d), $\alpha=1.005$ . In panel (a), the solid line is the plot of Eq. (\ref{ant}) with $\alpha=1.99$ and the dashed line represents the best fit for the temperature profile at $N=512$ with the heuristic formula (\ref{teo23}), the result of the fit is $\alpha=1.99$. In panel (b), we set $\alpha=1.00$ and it is clear that the temperature profile approaches the asymptotic temperature profile ($\ref{ant}$) as the length of the channel is increased.}
\end{figure}

The behavior of the temperature profile for four different cases is shown in Fig. \ref{figu5}. It is clear that in case I the temperature profiles in the central part of the channel are linear and maintain  a similar  shape for different system sizes. This behavior is quite similar to that presented in references \cite{tria,chaos}, when the  first system have no disorder and the second one exhibits regular dynamics. There are jump discontinuities in the temperature profile for cases I and II at the ends  of the channel. This phenomenon is the result of boundary heat resistance that usually appears when there is a heat flux across the interface of two adjacent materials (see. Ref. \cite{lepri} and references therein).  These jumps are finite size effects because as the system size increases boundary heat resistance decreases, as do  the temperature jumps. This is seen clearly in Fig. \ref{figu5}(b). When  $\beta \approx 0$ there are almost no jumps (see Figs. \ref{figu5}(c) and \ref{figu5}(d))  and the temperature profile can be approximated by the formula  (see Ref. \cite{chaos} for details):
\begin{equation}
T(x)=\frac{T_LT^{\frac{\alpha}{2}}_R(1-x)^{\gamma}+T_RT^{\frac{\alpha}{2}}_Lx^{\gamma}}{T^{\frac{\alpha}{2}}_R(1-x)^{\gamma}+T^{\frac{\alpha}{2}}_Lx^{\gamma}},
\label{ant}
\end{equation}
which is valid for large system sizes. Here $x=\frac{i}{N}$, $\gamma= (2-\alpha)\alpha^{\frac{3}{2}}$ and $\alpha$ is the diffusion exponent (keeping in mind that the diffusion exponent is not a constant, though it can be treated as such for relative long periods of time). For case I,  the temperature profile predicted by  (\ref{ant}) is clearly different from our numerical simulation. The numerical simulation and temperature profile predicted by this formula have a similar shape but are displaced (see Fig. \ref{figu5}(a)). However, we can see  that the heuristic formula
\begin{equation}
T(x)=\frac{T_LT^{\frac{1}{2}}_R(1-x)^{\gamma}+T_RT^{\frac{1}{2}}_Lx^{\gamma}}{T^{\frac{1}{2}}_R(1-x)^{\gamma}+T^{\frac{1}{2}}_Lx^{\gamma}}
\label{teo23}
\end{equation}
shows good agreement with the numerical results for case I (see dashed line in Fig. \ref{figu5}(a)). Even in the case of the flat channel ($\alpha=2$), there is a discrepancy between formula (\ref{ant}) and (\ref{teo23}), since the former predicts a flat profile given by $T(x)=\frac{2T_RT_L}{T_R+T_L}$, while formula (\ref{teo23}) also predicts a flat temperature profile with the value $T(x)=\sqrt{T_RT_L}$, but the latter flat temperature profile is an analytical exact result that can be obtained directly from definition (\ref{tem}). When the formula (\ref{ant}) was derived, a definition of temperature in terms of the mean density of particles at the cell $C_i$ was used. Nevertheless, definition (\ref{tem}) is the time average of kinetic energy at cell $C_i$; therefore, these two definitions lead to  different temperature profiles that are only equal when diffusion is normal. This explains the clear discrepancy between our numerical simulations and Eq. (\ref{ant}) for case I. In any case, both definitions of temperature lead to temperature profiles that are closely related to diffusion exponent $\alpha$ for sufficiently large system sizes. If a system obeys Fourier's law ($\alpha=1$), then the formula derived in Ref. \cite{lorentz} is recovered. In addition, when the temperature difference of the reservoirs is small,  the typical linear temperature profile will be obtained.

\section{Conclusions}
We analyzed the thermal transport properties of a sinusoidal channel placed between two Gaussian type thermal baths. The dynamics of a particle moving in the corresponding  infinite length channel exhibits Poincar\'e plots with a KAM structure typical of generic Hamiltonian systems.  When the ripple amplitude is small, the dynamics of the system is regular. For this case, estimates of the heat flux and temperature profile were obtained using an effective potential. Specifically, the temperature profile of the central part of the channel is well approximated by a flat profile given by the geometric average of the temperatures of the two reservoirs. Consequently, the same result is to be expected for other billiard systems with similar dynamics. In the regime of regular dynamics, the thermal conductivity  scales with the system size  as $\kappa \sim N$ for sufficiently large system size and the diffusion of particles is ballistic.  Unidirectional mixed chaos appears as the ripple amplitude is increased; and the mean square displacement can be approximated by $Dt^\alpha$ (being $\alpha \lesssim 2$ a constant) for a relative long time period. When bidirectional mixed chaos appears, the mean square displacement grows asymptotically as $t\log\left( t\right)$, then diffusion exponent is time dependent, but it remains practically constant for  relative long time periods. Temperature profiles were also analyzed for different degrees of chaos in the system, and it was found that the diffusion exponent is closely related to the temperature profile of the system.

\acknowledgments
The authors  thank to CONACYT, M\'exico and CIC-UMSNH for financial support.

\end{document}